# Reply to the comment on the paper "Thermodynamics of two-dimensional magneto nanoparticles (P. Vargas, D. Altbir, M.Knobel and D. Laroze)" by H. Büttner and Yu. Gaididei.

P. Vargas[1], D. Altbir[2], M. Knobel[3] and D. Laroze[4]

[1] *Departamento de Física, Universidad Técnica Federico Santa María Valparaíso, Chile*
[2] *Departamento de Física, Universidad de Santiago de Chile - Av. Ecuador 3493, Santiago, Chile*
[3] *Instituto de Física Gleb Wataghin, Universidad Estadual de Campinas (UNICAMP) Caixa Postal 6165, Campinas 13083-970, SP, Brasil*
[4] *Instituto de Física, Universidad Católica de Valparaíso - Valparaíso, Chile*

It is shown that there is bi-stability in a two dimensional system consisting of non interacting magnetic nanoparticles with equal uniaxial anisotropies. It is also shown that bi-stability still remains in three dimensions. The only consideration is that the applied magnetic field has to be perpendicular to the anisotropy axis.

Recently appeared a comment [1] concerning our paper "Thermodynamics of two-dimensional magnetic nanoparticles". [2]
First of all, we have to clearly stated that our system is an ensemble of non-interacting magnetic nanoparticles living in a strict 2D world, i.e. all of them are located in a plane and with the same anisotropy easy axis. Moreover the magnetization of each particle is restricted to move in a plane, $\vec{m} = (\sin\theta, \cos\theta)$, i.e. $0 < \theta < 2\pi$ .and the magnetic field perpendicular to the easy axes also lies in the same plane. Mathematically in our model the magnetization has only one degree of freedom.
Experimentally using evaporation or lithography techniques one can fabricate a number of elongated magnetic nanoparticles (Ni) on flat surfaces (Cu, Pt) [3]. Therefore the particular configuration of our letter can be achieved.
Then, the comment[1] lack of physical significance.

Moreover, even if one considers a real 3D system, then Eqs. 4 and 5 of the comment [1] are misleading because the azimuthal factor is missing.
Usually, the energy of a non-interacting magnetic particle is given by (Zeeman and anisotropy energy terms):

$$E = -\vec{m}\cdot\vec{H} - \frac{1}{2}mH_a\left(\frac{\vec{m}\cdot\hat{n}}{m}\right)^2$$

In the particular case of external field along the z axis, $\vec{H} = (0,0,H)$, and anisotropy axis along the x axis, $\hat{n} = (1,0,0)$ one has that:

$$E = -mH\cos\theta - \frac{1}{2}mH_a\sin^2\theta\cos^2\phi \;,$$

being $\phi$ and $\theta$ the azimuthal and polar angle of the magnetization vector, $\vec{m} = (\sin\theta\cos\phi, \sin\theta\cos\phi, \cos\theta)$. Therefore in this particular case the partition function reads.

$$Z = \int_0^{2\pi} d\phi \int_0^{\pi} d\theta \sin\theta \exp\left(\frac{mH}{kT}\cos\theta + \frac{mH_a}{2kT}\sin^2\theta\cos^2\phi\right)$$

Therefore, the 3D case contains the factor $\cos^2 \phi$ which is missing in Eqs. 4 and 5 of the comment. One could argue that the 2D case is recovered by doing $\phi = 0$ in the last integral, but this is physically wrong because then the magnetization is not allowed to rotate in the full XZ-plane. In such case the polar angle covers only half of the XZ-plane ($0 < \theta < \pi$).

The last integral can be more conveniently solved by choosing the *z*-axis as the anisotropy easy axis, and the external field along the *x*-axis. In that case the above integral reduces to:

$$Z = \frac{1}{4\pi} \int_0^{2\pi} d\phi \int_0^{\pi} d\theta \sin\theta \exp\left(\frac{mH}{kT}\sin\theta\cos\phi + \frac{mH_a}{2kT}\cos^2\theta\right)$$
$$= \frac{1}{2}\int_0^{\pi} d\theta \sin\theta \exp\left(\frac{mH_a}{2kT}\cos^2\theta\right) I_0\left(\frac{mH}{kT}\sin\theta\right) \quad (1)$$

Therefore the magnetization along the magnetic field is given by the following expression

$$M = m\frac{\partial \ln Z}{\partial \xi}, \quad \text{where } \xi = \frac{mH}{kT}.$$

Figure 1 depicts the result of the magnetization (in Bohr magnetons) as a function of Temperature (in Kelvin), evaluated using the Integral (1), with the following parameters, $m = 1000\mu_B$, $H_a = 3kOe$, and $H = 0.1kOe$ [3].

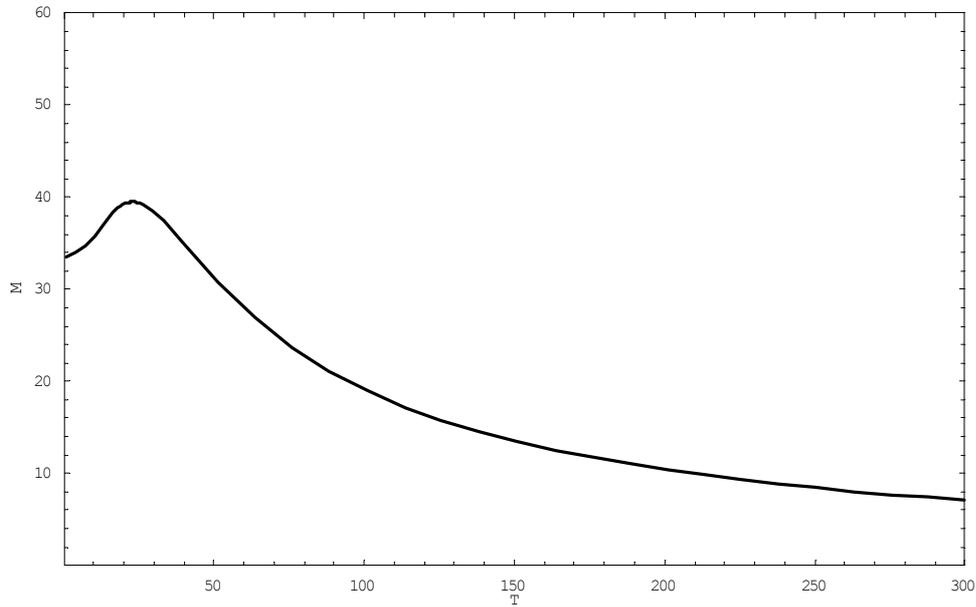

Fig.1 Magnetization along the external field (units of $\mu_B$) as a function of Temperature (in Kelvin) for a free 3D magnetic nanoparticle with $m = 1000\mu_B$. External field is *H*=3kOe and perpendicular to the easy axis. The anisotropy field is of $H_a$=3kOe. (compare with Fig. 4 of Ref [2], the 2D case)

Concluding, we demonstrated that even in the 3D case, a free magnetic particle with uniaxial anisotropy has a magnetization maximum at finite temperature when the external field points perpendicular to the easy axis.

Although the physics behind this case is the same, the 2D case, which was studied in our letter, is simpler because allows an easier analytical expression for the partition function and has a straight similarity with the mechanical rotating system.

[1]H. Büttner, Yu. Gaididei, arXiv:cond-mat/0304156 v1 7Apr 2003

[2] P. Vargas, D.Altbir, M.Knobel and D. Laroze, Europhys. Lett. **58**,603 (2002) (available in the web page http://www.nucleo-milenio.cl )

[3]Chen J.Y, Ross C.A., Thomas E.L. Lammertink R.G.H., Vancso G.J., IEEE T. Magn. 38(5), 2541 (2002)

[4] P. Vargas and D. Laroze, submitted to JMMM (2003)